\documentclass[12pt]{iopart}
\usepackage{graphicx}
\usepackage{amssymb,color}
\usepackage{cite}
\pdfoutput=1

\def\abs#1{\left|#1\right|}

\begin{document}
\title[Generalised diamond chain model for azurite]{Dynamic and
thermodynamic properties of the generalised
diamond chain model for azurite}
\author{Andreas Honecker$^1$, Shijie Hu$^1$,
Robert Peters$^2$ and Johannes Richter$^3$}
\address{$^1$ Institut f\"ur Theoretische Physik,
 Georg-August-Universit\"at G\"ottingen, 37077 G\"ottingen, Germany}
\address{$^2$ Department of Physics, Graduate School of Science,
 Kyoto University, Kyoto 606-8502, Japan}
\address{$^3$ Institut f\"ur Theoretische Physik,
 Otto-von-Guericke-Universit\"{a}t Magdeburg,
 P.O.\ Box 4120, 39016 Magdeburg, Germany}
\ead{ahoneck@uni-goettingen.de}


\begin{abstract}
The natural mineral azurite Cu$_3$(CO$_3$)$_2$(OH)$_2$ is an interesting
spin-1/2 quantum antiferromagnet.
Recently, a generalised diamond chain model has
been established as a good description of the magnetic properties of
azurite with parameters
placing it in a highly frustrated parameter regime.
Here we explore further properties of this model for azurite. First,
we determine the inelastic neutron scattering spectrum in the absence
of a magnetic field and find good agreement with experiments, thus
lending further support to the model. Furthermore, we present numerical
data for the magnetocaloric effect and predict that strong cooling
should be observed during adiabatic (de)magnetisation of azurite
in magnetic fields slightly above $30$T. Finally,
the presence of a dominant dimer interaction in azurite suggests the
use of effective Hamiltonians for an effective low-energy description
and we propose that such an approach may be useful to fully account
for the three-dimensional coupling geometry.
\end{abstract}
\pacs{
75.10.Jm, 
75.30.Sg, 
02.70.-c, 
78.70.Nx  
}
\vspace{28pt plus 10pt minus 18pt}
     \noindent{\small\rm To appear in: {\it \JPCM}\par}
\maketitle

\section{Introduction}

On the one hand, highly frustrated magnets constitute a fascinating field
of research since the competition of different interactions give rise to
many exotic phenomena (see for example \cite{HFMbook}). On the other hand,
theoretical studies of highly frustrated quantum magnets are usually a
notoriously difficult task, for example because the so-called
`sign problem' precludes efficient Quantum Monte Carlo simulations of
such models \cite{TrWi05}. A notable exception to this general rule
are models which allow for the construction of exact ground states
because of destructive quantum interference caused exactly by the
frustrating interactions. A famous example is the exact dimer
ground state of the two-dimensional
Shastry-Sutherland model for SrCu$_2$(BO$_3$)$_2$ (see \cite{MiUe03}
for a review).

Another case of such exact eigenstates are the ground states which
can be constructed exactly in terms of localised magnons in the
high-field regime of certain highly frustrated quantum magnets
\cite{SSRS01,loc_mag02,LNP04,ZhHo04,ZhiTsu04,DR04,RSHS04,SP2004,loc_mag04,ZhiTsu05,SRS05,%
DeRi06,SSHSR06,SRS07,ZT07,DRHS07,RLM08,RDH08,Schnack10}.
Remarkably, these localised ground states
give rise to a macroscopic degeneracy, i.e.\ a finite zero-temperature entropy
exactly at the saturation field 
\cite{LNP04,ZhHo04,ZhiTsu04,DR04,loc_mag04,ZhiTsu05,DeRi06,SSHSR06,ZT07,DRHS07,RDH08}.
This implies an enhanced magnetocaloric effect and promises
applications for efficient low-temperature cooling
\cite{ZhHo04,SRS07,Zhito03,SPSGBPBZ05,DeRi06,ZhHo09,PML09}.

\begin{figure}[t]
\begin{center}
\includegraphics[width=0.5\columnwidth]{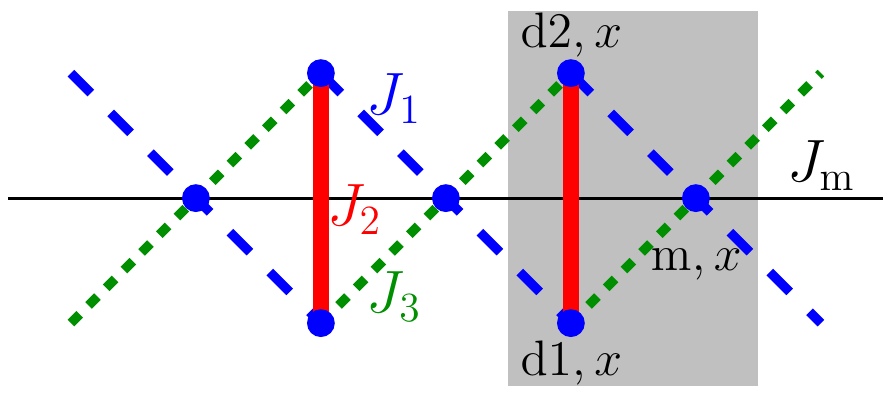}
\end{center}
\caption{
    Generalised diamond chain model. A unit cell $x$ (indicated by the
    grey shaded region) contains two dimer sites `${\rm d1},x$,' `${\rm d2},x$'
    and one monomer site `${\rm m},x$'. These sites are
    connected by the exchange constants $J_1$, $J_2$, $J_3$ and
    $J_{\rm m}$ which are indicated by different line styles.
}
\label{fig:model}
\end{figure}

Models with local conservation laws
\cite{DR04,DeRi06,DRHS07,Gelfand91,IvRi97,RIS98,MTM99,KOK00,HMT00,MHSKU00,%
GVAHW00,TrSe00,HoBr01,ChBo02,SchRi02a,SchRi02b,ChBue02,RoAl03,RDT06,BWB08,MPSM10,%
DTK10}
can be considered as a special mechanism to ensure
the presence of localised magnons. One particular
model in this category is the {\it ideal diamond chain} whose ground-state
phase diagram was studied in \cite{TKS96}. This model is sketched
in figure \ref{fig:model}; the ideal diamond chain is obtained by
setting $J_1=J_3$ and $J_{\rm m}=0$. In the case $J_1=J_3$, the total
spin $\vec{S}_{{\rm d1},x} + \vec{S}_{{\rm d2},x}$
of a vertical dimer is a conserved quantity  in each unit cell $x$.
Several modifications of the ideal diamond chain have also   
been considered in the recent  literature, see, e.g.
\cite{strechka1,strechka2,ivanov,hida1,hida2,rsok10}.
In particular, the `{\it distorted}' variant of the spin-1/2
diamond chain with $J_1 \ne J_3$ and $J_{\rm m}=0$
has attracted much attention from the theoretical side
\cite{OTTK99,TOHTK00,TOHTK01,HoL,OTK03,SOT09}. Among the theoretical results
we would like to mention in particular that a plateau at one third
of the saturation magnetisation is abundant in the spin-1/2 distorted diamond
chain, as can be expected for a model with a unit cell of three sites
\cite{OYA97,CHP97,CHP98}. Evidently, it is very desirable to have an
experimental realisation of a spin-1/2 distorted diamond chain,
preferably in the highly frustrated regime $J_1 \approx J_3$.

The natural mineral azurite Cu$_3$(CO$_3$)$_2$(OH)$_2$
was originally suggested to realise a spin-1/2 distorted diamond chain
with all exchange constants antiferromagnetic \cite{KikuchiA,KikuchiB,MiLu},
i.e.\ $J_1$, $J_2$, $J_3 > 0$ and $J_{\rm m}=0$ in figure \ref{fig:model}.
This picture had, however, been questioned: Some authors have suggested a
ferromagnetic $J_3<0$ \cite{GuSu1,GuSu2,azuritINS} which would render the
model non-frustrated whereas other authors have argued interchain
coupling to be important \cite{Whangbo}. Recent first-principles
density-functional computations \cite{JeschkeEtal} indeed yield a
three-dimensional coupling geometry with a dominant antiferromagnetic
dimer exchange constant $J_2 > 0$. Nevertheless, closer inspection of
the exchange geometry allows one to map this three-dimensional network
effectively to the {\it generalised diamond chain}
sketched in figure \ref{fig:model} \cite{JeschkeEtal}.
A small refinement of the exchange constants obtained from the first-principles
density-functional computations led to \cite{JeschkeEtal}
\begin{equation}
 J_1=15.51 {\rm K}\, , \qquad
 J_2=33 {\rm K} \, , \qquad
 J_3=6.93 {\rm K} \, , \qquad
 J_{\rm m} = 4.62 {\rm K} \, .
 \label{eq:Jazurite}
\end{equation}
Using different variants of the density-matrix renormalisation
group (DMRG) method \cite{DMRGa,DMRGb}, it
was demonstrated \cite{JeschkeEtal}
that the generalised diamond chain with the values (\ref{eq:Jazurite})
of the exchange constants is consistent with a broad range of
experiments, namely
the magnetisation curve \cite{KikuchiA,KikuchiE},
the magnetic susceptibility \cite{KikuchiA,JeschkeEtal},
the specific heat \cite{KikuchiA,azuritINS,JeschkeEtal},
the structure of the the one-third plateau as determined by NMR \cite{azuritNMR}
and last but not least inelastic neutron scattering on this
one-third plateau \cite{azuritINS}.

It should be noted that the parameter set (\ref{eq:Jazurite}) is not
very far from the original proposal \cite{KikuchiA}. In particular,
the fact that the two exchange constants $J_1$ and $J_3$ are of
a comparable magnitude places azurite in a highly frustrated parameter
regime. The main difference between the original
parameter set \cite{KikuchiA} and (\ref{eq:Jazurite}) is
the direct exchange coupling  $J_{\rm m}$ between
monomer spins whose presence was already suggested in \cite{azuritINS}.

In order to be precise, we present the Hamiltonian for the
generalised diamond chain sketched in figure \ref{fig:model}:
\begin{eqnarray}
{\cal H} &=&
\sum_{x=1}^{N/3} \left\{
J_1 \, \vec{S}_{{\rm m},x} \cdot
   \left(\vec{S}_{{\rm d2},x} + \vec{S}_{{\rm d1},x+1} \right)
+ J_2 \, \vec{S}_{{\rm d1},x} \cdot \vec{S}_{{\rm d2},x}
\right.
\nonumber\\
&& \left. \qquad
+ J_3 \, \vec{S}_{{\rm m},x} \cdot
   \left(\vec{S}_{{\rm d1},x} + \vec{S}_{{\rm d2},x+1} \right)
+ J_{\rm m} \, \vec{S}_{{\rm m},x} \cdot \vec{S}_{{\rm m},x+1}
\right\}
\nonumber\\
&&
- g\,\mu_B\,H\, \sum_{x=1}^{N/3} \left(S_{{\rm d1},x}^{\rm z}
+ S_{{\rm d2},x}^{\rm z}
+ S_{{\rm m},x}^{\rm z}\right) \, .
\label{eq:HgenDiamondChain}
\end{eqnarray}
The total number of spins is denoted by $N$ and $x$ runs over the
$N/3$ unit cells.
The $\vec{S}_{\cdot,x}$ are spin-1/2 operators, $H$ the external magnetic
field and $\mu_B$ the Bohr magneton. In order to express the magnetic
field in experimental units, we need the value of the gyromagnetic ratio
$g$. Here we follow \cite{JeschkeEtal} and use $g = 2.06$ which is
consistent with high-field ESR on azurite \cite{Ohta}.

This paper is organised as follows. In section \ref{sec:effH}
we discuss two effective Hamiltonians which are obtained \cite{HoL} by
applying strong-coupling perturbation theory to the model
(\ref{eq:HgenDiamondChain}).
The two effective Hamiltonians describe the low-energy
(low-temperature) behaviour in the regime of magnetisation up to
one-third and between one-third and full magnetisation, respectively.
In section \ref{sec:SpecH0} we then compute the zero-field excitation
spectrum by exact diagonalisation and a dynamical variant of the
DMRG method \cite{dDMRG}. We observe good agreement with the inelastic
neutron scattering results \cite{azuritINS}, thus lending further support to the
description of azurite in terms of the generalised diamond chain
model (\ref{eq:HgenDiamondChain}) with the parameters (\ref{eq:Jazurite}).
Next, we explore magnetocaloric properties of the model in section
\ref{sec:Mcal} using computations based on a transfer-matrix variant of the
DMRG method \cite{TMRG96,TMRG97}. The zero-temperature entropy which is
present in the ideal diamond chain exactly at the saturation field
\cite{DRHS07,DeRi06} is lifted by the distortion $J_1 \ne J_3$.
Nevertheless, we predict that cooling down to temperatures
substantially below $1$K should be possible in the high-field regime.
Finally, in
section \ref{sec:SumDisc} we summarise our findings and suggest
topics for further theoretical and experimental studies of azurite.

\section{Low-energy effective Hamiltonians}

\label{sec:effH}

A simple picture of azurite is given by
an effective spin-1/2 Heisenberg chain accounting for the low-energy
excitations at small magnetic fields and weakly
coupled dimers which describe higher energies or
higher magnetic fields
\cite{KikuchiA,azuritINS,MiLu,Whangbo,JeschkeEtal}.
The corresponding effective Hamiltonians can easily
be obtained from the results of \cite{HoL,JeschkeEtal}.
We will nevertheless discuss them here since they will be
useful for the later analysis.

\subsection{Small magnetic fields}

\label{sec:effSmallH}

At small magnetic fields and for large $J_2$ the dimers are frozen
in their singlet ground state. Accordingly, the low-energy degrees of
freedom are given by the monomer spins in this low-field regime.
For the generalised diamond chain model, the monomer spin degrees
give rise to an effective spin-1/2 chain. This 
description holds up to the one-third plateau where all spins of the
effective spin-1/2 chain are aligned along the field direction.
Because of the SU(2) symmetry
present at $H=0$, this effective spin-1/2 chain can
contain only SU(2) symmetric terms. Considering the limit $J_2 \gg
\abs{J_i}$ ($i =1,3$) \cite{HoL} and $H=0$
one arrives at a nearest-neighbour Heisenberg chain
\begin{equation}
{\cal H}_{\rm eff.}^{\rm m} = J_{\rm eff.}
\sum_{x} \vec{S}_{{\rm m},x} \cdot \vec{S}_{{\rm m},x+1} \, .
\label{eq:HeffM}
\end{equation}
The effective exchange constant $J_{\rm eff.}$ is given
up to second order in $\abs{J_i} \ll  J_2$ ($i =1,3$) by \cite{HoL}
\begin{equation}
J_{\rm eff.} = J_{\rm m} + \frac{(J_1-J_3)^2}{2\,J_2} +
{\cal O}\left(\left\{\frac{J_i^3}{J_2^2}\right\}\right) \, .
\label{eq:JeffPert}
\end{equation}
Insertion of the values (\ref{eq:Jazurite}) for the exchange constants
yields $J_{\rm eff.} \approx 5.8$K.

\begin{figure}[t]
\begin{center}
\includegraphics[width=0.7\columnwidth]{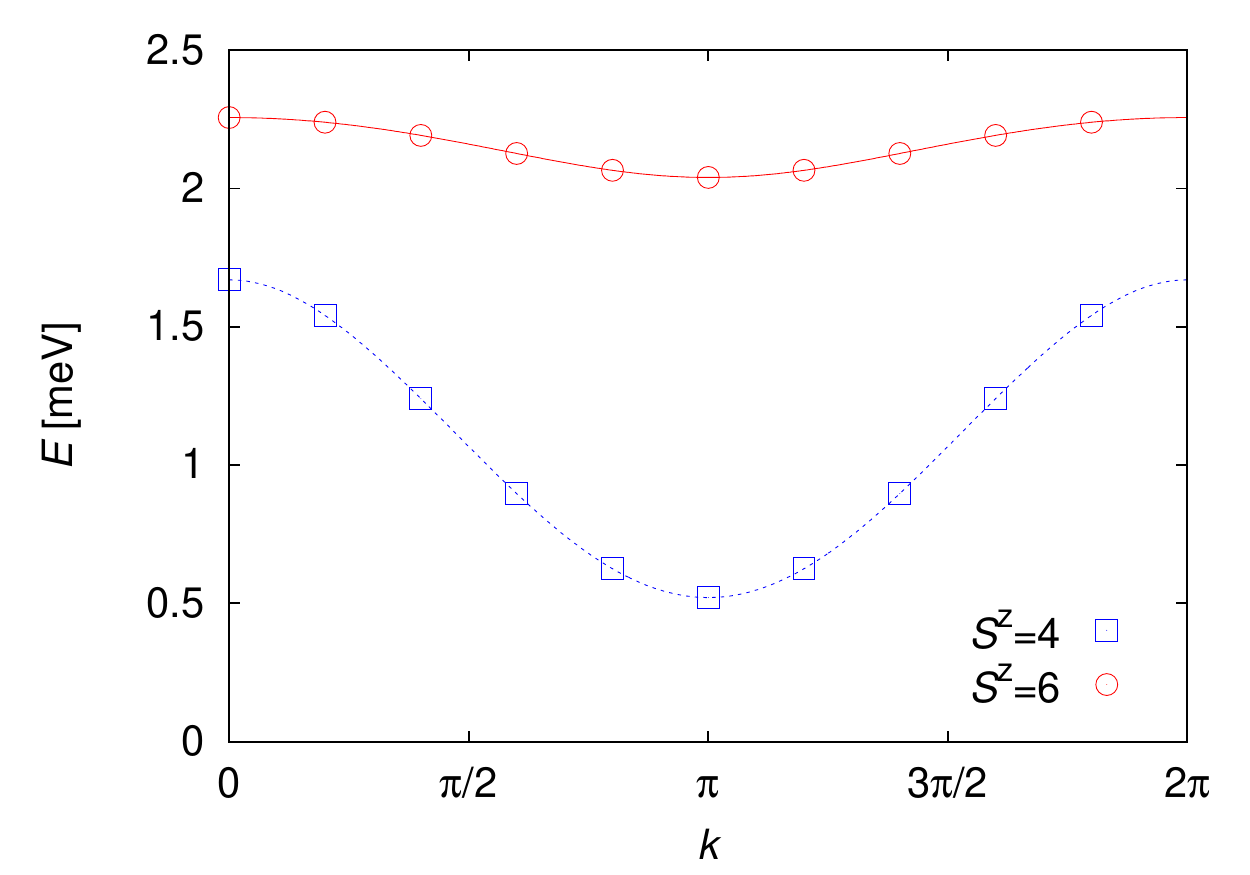}
\end{center}
\caption{
    Excitations of the generalised diamond chain with
    $N=30$ sites on the one-third plateau
    obtained by exact diagonalisation with
    $J_1=15.51$K, $J_2=33$K, $J_3=6.93$K, $J_{\rm m} = 4.62$K
    and $H=14$T. Squares show results for $S^{\rm z} = 4$
    which is just below the plateau and circles show results for
    $S^{\rm z} = 6$ which is just above the plateau value
    $S^{\rm z} = N/6 = 5$. Lines show interpolations which have
    been obtained by a Fourier analysis.
}
\label{fig:plateauSpecN30}
\end{figure}

A more accurate estimate for $J_{\rm eff.}$
can be obtained from an analysis of the spectrum
on the one-third plateau where one finds two sharp
excitation branches: the lower one can be attributed to the effective
spin-1/2 chain and the upper one to the dimers. The numerical results
\cite{JeschkeEtal} for these two branches at $N=30$ are reproduced
in figure \ref{fig:plateauSpecN30}. Lines in this figures have
been obtained from a Fourier analysis of the finite-size data.
{}From the first Fourier component of the lower branch one
finds the effective exchange constant
\begin{equation}
J_{\rm eff.} = 6.595 {\rm K} 
\label{eq:valJeff}
\end{equation}
for the parameter set (\ref{eq:Jazurite}).
Higher harmonics contribute less than $4\%$ of the first harmonic
to the dispersion of this lower branch, consistent with only
nearest-neighbour interactions appearing in the effective
Hamiltonian (\ref{eq:HeffM}).

A consistency check on this picture is obtained by the lower
edge of the one-third plateau which is located at
$H_{c1} = 13.32$K 
according to numerical data for
the generalised diamond chain model with the
parameters (\ref{eq:Jazurite}) \cite{JeschkeEtal}.
The transition at $H_{c1}$ corresponds to the transition
to saturation of the effective spin chain (\ref{eq:HeffM}). This predicts
the equality  $H_{c1} = 2\,J_{\rm eff.}$ which indeed holds
to high accuracy with (\ref{eq:valJeff}).

\subsection{High magnetic fields}

A similar effective description
holds at high magnetic fields. If $J_2$ is dominant
and $H \approx J_2$, the monomer spins are fully polarised whereas
the singlet state
$\vert {\rm s} \rangle = \frac{1}{\sqrt{2}} \, \left(
\vert \uparrow \downarrow \rangle - \vert \downarrow \uparrow \rangle
\right)$
and the spin polarised component of the triplet
$\vert {\rm t} \rangle = \vert \uparrow \uparrow \rangle$
are (almost) degenerate. The collective behaviour of these dimer
degrees of freedom can be efficiently encoded by pseudo-spin-1/2
operators acting at site $x$
\begin{eqnarray}
T^{\rm z}_x \, \vert {\rm s} \rangle_x &=& -\frac{1}{2} \, \vert {\rm s} \rangle_x
\, , \qquad
T^{\rm z}_x \, \vert {\rm t} \rangle_x = \frac{1}{2} \, \vert {\rm t} \rangle_x \, ,
\nonumber \\
T^+_x \, \vert {\rm s} \rangle_x &=& \vert {\rm t} \rangle_x \, , \qquad
T^+_x \, \vert {\rm t} \rangle_x = 0 \, ,
\nonumber \\
T^-_x \, \vert {\rm t} \rangle_x &=& \vert {\rm s} \rangle_x \, , \qquad
T^-_x \, \vert {\rm s} \rangle_x = 0 \, .
\label{eq:defT}
\end{eqnarray}
The dimer degrees of freedom give rise to another effective spin
chain. However, this effective spin chain is anisotropic since the
presence of an external magnetic field reduces the symmetry to U(1).
In addition, corrections to the magnetic field $H$ also need to be
taken into account. Hence, the effective Hamiltonian for the $N/3$
dimer degrees of freedom is
\begin{eqnarray}
{\cal H}_{\rm eff.}^{\rm d} &=&
\sum_{x=1}^{N/3} \left\{J_{\rm z} \, T^{\rm z}_{x} \, T^{\rm z}_{x+1}
+ \frac{J_{\rm xy}}{2} \left(T^+_{x} \, T^-_{x+1}
+ T^-_{x} \, T^+_{x+1} \right) \right\}
\nonumber\\
&& - (H-J_{\rm dimer}) \sum_{x} T^{\rm z}_x  \, .
\label{eq:HeffD}
\end{eqnarray}
The effective exchange constants can again be determined by
second-order perturbation theory
in $\abs{J_i} \ll  J_2$ ($i =1,3$) \cite{HoL}:
\begin{eqnarray}
J_{\rm xy} &=& \frac{(J_1-J_3)^2}{4\,J_2} +
{\cal O}\left(\left\{\frac{J_i^3}{J_2^2}\right\}\right) \, ,
\nonumber\\
J_{\rm z} &=& 
{\cal O}\left(\left\{\frac{J_i^3}{J_2^2}\right\}\right) \, ,
\nonumber\\
J_{\rm dimer}  &=&  J_2 + \frac{J_1+J_3}{2}
+ \frac{(J_1-J_3)^2}{4\,J_2} +
{\cal O}\left(\left\{\frac{J_i^3}{J_2^2}\right\}\right) \, .
\label{eq:JDpert}
\end{eqnarray}
In passing we note that from (\ref{eq:JeffPert}) and
(\ref{eq:JDpert}) one finds $J_{\rm xy}/J_{\rm eff.} = 1/2$
for $J_{\rm m} = 0$ and up to second order in $J_1$ and $J_3$.
This ratio translates directly into the ratios of the bandwidths
of the two excitation branches on the one-third plateau. However,
in azurite this bandwidth ratio is measured to be about 1/6
\cite{azuritINS}. This indicates that the excitation spectrum
of azurite on the one-third plateau \cite{azuritINS} cannot
be fitted by a simple distorted diamond chain with $J_{\rm m}=0$
\cite{JeschkeEtal}, at least not in the region of large $J_2$.

Insertion of (\ref{eq:Jazurite}) into (\ref{eq:JDpert}) yields
$J_{\rm xy}  = 0.56$K, 
$J_{\rm z} = 0$ and
$J_{\rm dimer} = 44.8$K. 
More precise values for the effective parameters can again be derived
by an analysis of numerical data \cite{JeschkeEtal}
for the parameters (\ref{eq:Jazurite}). The first Fourier coefficient
of the upper branch in figure \ref{fig:plateauSpecN30} yields
\begin{equation}
J_{\rm xy}  = 1.249 {\rm K} \, . 
\label{eq:JxyVal}
\end{equation}
Higher Fourier components contribute less than $7\%$ of the
first component, in agreement with only nearest-neighbour
interactions appearing in (\ref{eq:HeffD}).

The other parameters can for instance be determined from the upper
edge $H_{c2}$ of the one-third plateau and the transition to saturation
at $H_{\rm sat.}$. Using
$H_{c2} = J_{\rm dimer} - J_{\rm xy} - J_{\rm z}$,
$H_{\rm sat.} = J_{\rm dimer} + J_{\rm xy} + J_{\rm z}$,
as appropriate for (\ref{eq:HeffD}),
the numerical values \cite{JeschkeEtal}
$H_{c2} = 43.045$K, 
$H_{\rm sat.} = 46.674$K 
and (\ref{eq:JxyVal}) we find
\begin{equation}
J_{\rm z} = 0.565 {\rm K} 
\, , \qquad
J_{\rm dimer} =  44.860{\rm K} 
\, . 
\label{eq:JzJdimerVal}
\end{equation}
Two remarks are in order at this point. Firstly, we note that
$J_{\rm dimer}$ differs substantially from the bare dimer coupling
constant $J_2$ which can be traced to substantial first-order
corrections in (\ref{eq:JDpert}). This precludes a direct
derivation of $J_2$ from most experimental data on azurite if
accurate results are desired. Secondly, we observe
that the numerical value $J_{\rm z} / J_{\rm xy} \approx 0.45$ lies in
the easy-plane regime of the effective Hamiltonian (\ref{eq:HeffD}).

A consistency check of the above analysis can be obtained from the
average position of the upper branch in figure \ref{fig:plateauSpecN30}
which yields
$J_{\rm dimer} - J_{\rm z} = 44.371$K. 
This agrees with (\ref{eq:JzJdimerVal}) to better than $0.1$K.

\section{Excitation spectrum in zero field}

\label{sec:SpecH0}

\begin{figure}[t]
\begin{center}
\includegraphics[height=0.64\columnwidth]{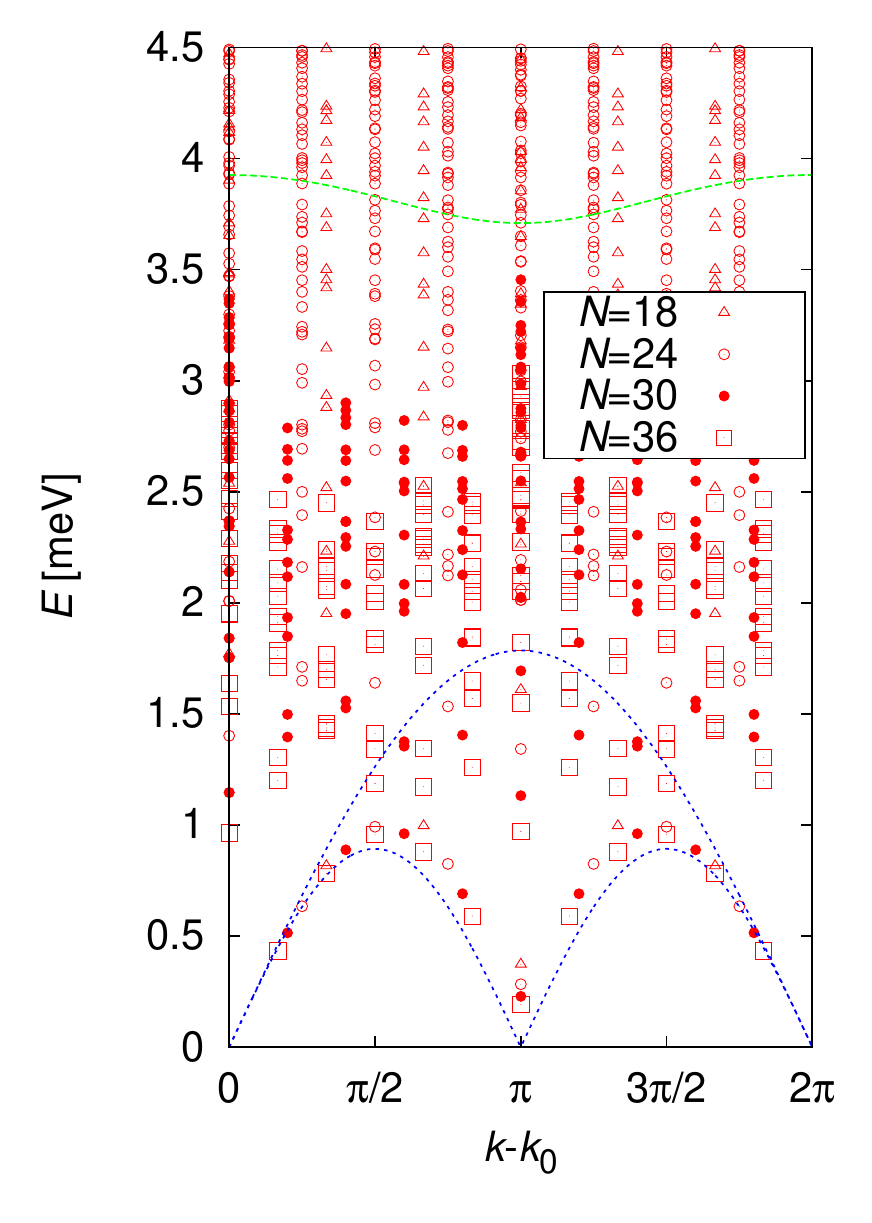}
\includegraphics[height=0.64\columnwidth]{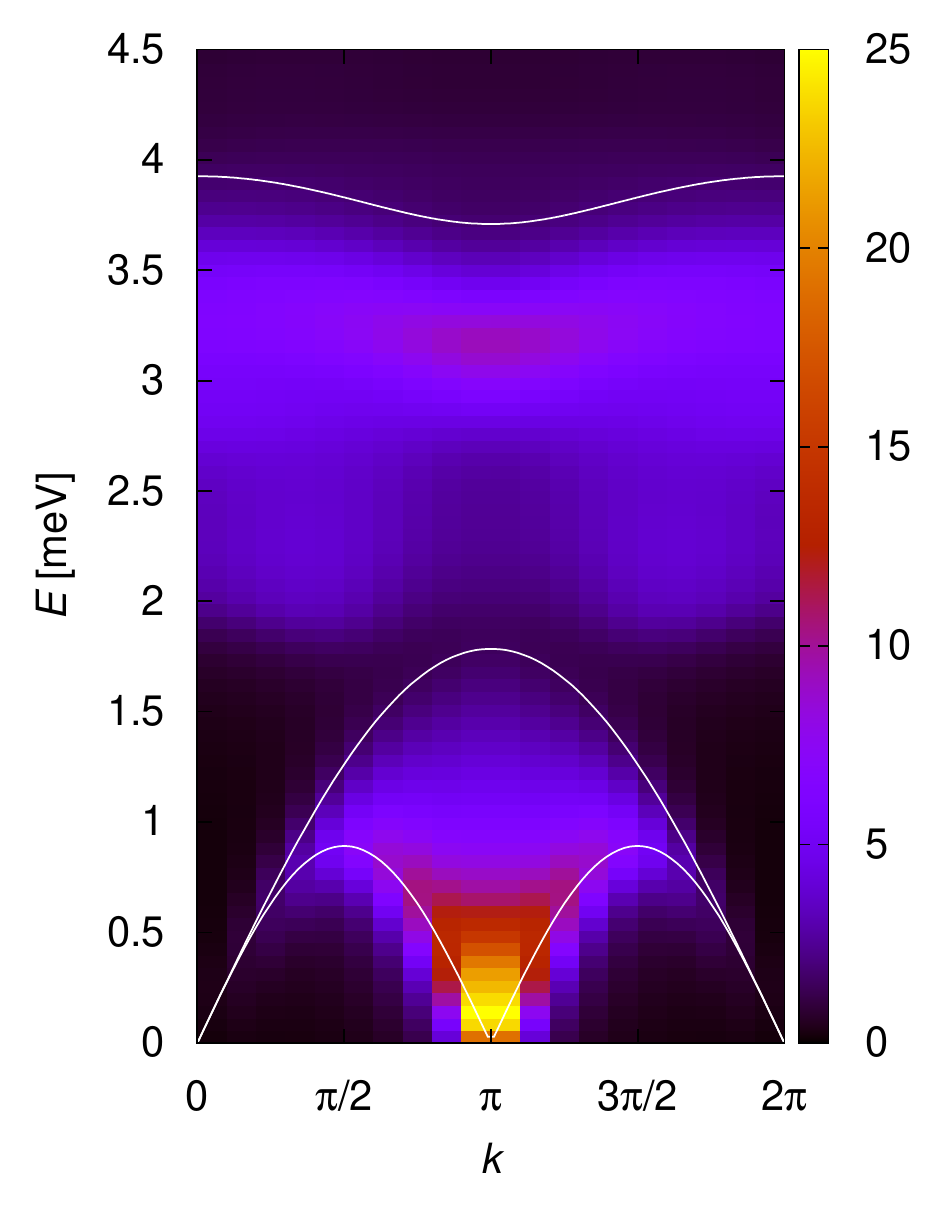}
\caption{
Zero-field spectrum of the generalised diamond chain model
for azurite with the exchange constants (\ref{eq:Jazurite}).
The left panel shows excitation energies $E$ in the spin-1 sector
obtained by exact diagonalisation for rings with $N=18$, $24$, $30$
and $36$ spins as a function of momentum $k-k_0$ where $k_0$
is the momentum of the ground state. The right panel shows
dynamical DMRG results for the dynamic structure factor of a system
with open boundary conditions and $N=60$ sites as a function
of energy $E$ and momentum transfer $k$. In the latter panel,
the shading corresponds to the neutron scattering
intensity in arbitrary units. \\
In both panels lines at low energies denote the boundaries of the two-spinon
continuum of an effective spin-1/2 Heisenberg chain with
an effective exchange constant $J_{\rm eff.} \approx 6.6{\rm K}
\approx 0.57$meV. The lines at an energy slightly below
$4$meV show the dimer excitation (upper branch) of figure
\ref{fig:plateauSpecN30} shifted up in energy by $H=14{\rm T} \approx 1.67$meV.
\label{fig:specH0}
}
\end{center}
\end{figure}

The excitation spectrum on the one-third plateau found by inelastic neutron
scattering on azurite \cite{azuritINS} has been
compared in \cite{JeschkeEtal}
to computations for the generalised diamond chain model
with the parameters (\ref{eq:Jazurite})
and very good agreement was found. Here we will perform a similar
comparison with the experimental excitation spectrum in
zero field \cite{azuritINS}.

First, we look at the spectrum itself which we have computed by
exact diagonalisation
data for periodic chains. In this case momentum $k$ is a good
quantum number. In addition, we use conservation of total $S^{\rm z}$
as well as spin-inversion for $S^{\rm z}=0$.
By virtue of the Wigner-Eckart theorem, inelastic neutron scattering on
the ground state of an SU(2) symmetric antiferromagnet is sensitive only
to excitations with total spin 1 \cite{MTBB81}.
Therefore
we reconstruct the total spin quantum numbers from the $S^{\rm z}$-
and spin-inversion-resolved
results. For $N=18$ spins it is possible to perform a full
diagonalisation. For $N=24$ sites we have used the method described in
section 2.1 of \cite{HoWe09} to compute a large number of low-lying
energies. Finally, for $N=30$ and $36$ we have used Spinpack
(see {\tt http://www-e.uni-magdeburg.de/jschulen/spin/index.html}).
The left panel of figure \ref{fig:specH0} shows the results of
these computations for the spin-1 sector.
For $N=18$ and $24$ we have sufficiently
many states to cover the shown energy range completely. However,
for $N=30$ and $36$ it is impossible to compute so many states
accurately. Therefore, for $N=30$ and $36$ we had to restrict to
lower energies $E \lesssim 2.8$meV and $2.4$meV, respectively.
Accordingly, one should keep in mind
that eigenvalues are missing in the left panel of figure \ref{fig:specH0}
at higher energies for the two biggest system sizes.

One can discern some structure
in the spectrum of the left panel of figure \ref{fig:specH0}
at low energies which we will discuss below.
However, at energies $E \gtrsim 1.5$meV, there is a large
density of spin-one excitations without any
evident structure.
In order to understand which excitations can be observed by
neutrons, we therefore need to compute the dynamic structure factor
\begin{equation}
S^{{\rm xx}}(\vec{k},\omega) = \frac{3}{N}\,
 \sum_{\langle x,y\rangle} {\rm e}^{i\,\vec{k}\cdot\left(\vec{R}_x-\vec{R}_y\right)}
\
{\rm Im} \langle 0 \vert
S^{\rm x}_x
\frac{1}{\omega+E_0-{\cal H}+i\,\eta}
S^{\rm x}_y
\vert 0 \rangle \, .
\label{eq:defSxx}
\end{equation}
We have computed this dynamic structure factor via the correction-vector
method in the density-matrix renormalisation group (DMRG) following \cite{dDMRG}. 
First, we have calculated the spectral function
between each pair of sites in the chain
\begin{equation}
        G_{x,y}(\omega)=\langle0\vert S_x^{\rm x}
\frac{1}{\omega+E_0-{\cal H}+i\,\eta}S_y^{\rm x}\vert0\rangle
\label{eq:defG}
\end{equation}
using $2$ sweeps for each
frequency point $\omega$, $N=60$ lattice sites and $m=200$ kept states.
For technical reason one has to introduce a Lorentzian broadening $\eta>0$
of the spectral function. We choose $\eta=0.05 \, J_2$.

In a second step, one needs to Fourier transform $G_{x,y}$ in order
to obtain the dynamic structure factor (\ref{eq:defSxx}).
The diamond chain contains dimers which are coupled strongly by $J_2$.
The structure factor of a dimer is known to depend strongly on the momentum
transfer and to lead to vanishing intensity at zero momentum transfer along
the dimer direction \cite{FuGue79}. It is therefore important to
use the precise positions $\vec{R}_x$ of the spin-1/2
copper atoms in azurite \cite{Zigan:72}
as well as the experimental value of the momentum transfer transverse to
the chains \cite{azuritINS} when evaluating (\ref{eq:defSxx}) from
$G_{x,y}$. The
result of this procedure is shown in the right panel of
figure \ref{fig:specH0}. The main limiting factor of the resolution
in this panel is a finite resolution of the momentum transfer $k$
along the chain direction which is caused by the open ends of the
chain.

The calculated structure factor shown in
the right panel of figure \ref{fig:specH0} shares the
following features with the results obtained by
inelastic neutron scattering on azurite \cite{azuritINS}:
(i) There is a scattering continuum at energies $E \lesssim 2$meV
with a sharp lower edge.
(ii) At higher energies, one finds a broad band of excitations with
a sharp feature in its middle. Just the energy of this feature differs
between the model where it is located at $E \approx 3.2$meV and the
experiment \cite{azuritINS} which finds it at $E \approx 6$meV.

For a more quantitative analysis of the low-energy features we
can use the effective spin-chain Hamiltonian introduced in
section \ref{sec:effSmallH}.
It is well known that the low-energy excitations of a spin-1/2 Heisenberg
chain form a two-spinon continuum whose boundaries are given by
\cite{MTBB81,ClPe62,FaTa81}
\begin{equation}
\epsilon_{\rm l} = \frac{\pi}{2}\,J_{\rm eff.}\,\abs{\sin{k}} \, ,
\qquad
\epsilon_{\rm u} = {\pi}\,J_{\rm eff.}\,\sin\frac{k}{2} \, .
\label{eq:2spinon}
\end{equation}
The boundaries of the two-spinon continuum with the value
of $J_{\rm eff.}$ given by (\ref{eq:valJeff}) are shown by the lines
in figure \ref{fig:specH0} and one observes that the low-energy
excitations fall indeed into this region. One can also directly compare
the lowest excitations in the left panel of figure \ref{fig:specH0} with
those of a spin-1/2 chain with $N/3$ sites \cite{MTBB81,ClPe62} and one observes
again good agreement. Furthermore, even the low-energy part of the
dynamic structure factor shown in the right panel of figure \ref{fig:specH0}
matches nicely with that of a spin-1/2 Heisenberg chain
\cite{LTFN05,CaHa06}. Hence we conclude that an effective
spin-1/2 Heisenberg chain with $J_{\rm eff.}$ given by (\ref{eq:valJeff})
describes the low-energy properties of the generalised diamond chain well.

The location of the dimer branch can be estimated by adding the
Zeeman energy $H=14{\rm T} \approx 1.67$meV to the upper branch in
figure \ref{fig:plateauSpecN30}. This yields the curves in
figure \ref{fig:specH0} slightly below $4$meV. The shape of
this curve traces the dispersion of the
broad maximum between $3$ and $3.5$meV
in the right panel of figure \ref{fig:specH0} nicely, but it is
about $0.5$meV too high in energy. Indeed, by applying a simple
Zeeman shift to the upper branch of figure \ref{fig:plateauSpecN30},
we are stretching the high-field effective Hamiltonian (\ref{eq:HeffD})
well beyond its limits of validity. After all, there are many low-energy
excitations at $H=0$ which are not taken into account by this Hamiltonian.
Renormalisation of the bare dimer excitation by
many-body effects is therefore expected.
This is reflected both by the broad diffusive
background and a down-shift of the maximum scattering intensity by
about $0.5$meV in the generalised diamond chain model
with respect to the effective Hamiltonian (\ref{eq:HeffD})
(see right panel of figure \ref{fig:specH0}).
It should be noted that the experiments \cite{azuritINS} on azurite
observe a renormalisation of the bare dimer excitation to higher energies
rather than lower energies as in our model. This is likely to be a signature
of the three-dimensional ordered state in which the experiments were
performed. Accordingly we speculate that quantitative agreement could
be improved by taking the three-dimensional coupling geometry \cite{JeschkeEtal}
fully into account.

In any case, analysis of the low-energy excitation spectrum of azurite
\cite{azuritINS} supports the conclusion \cite{JeschkeEtal} that the
generalised diamond chain model with the parameters (\ref{eq:Jazurite})
yields a good overall description of the experimental situation.

\section{Magnetocaloric effect in high magnetic fields}

\label{sec:Mcal}

One of the interesting features of azurite is its proximity to a class
of systems with localised magnons at high magnetic fields
giving rise to an enhanced magnetocaloric effect \cite{DRHS07}.

\begin{figure}[t]
\begin{center}
\includegraphics[width=0.7\columnwidth]{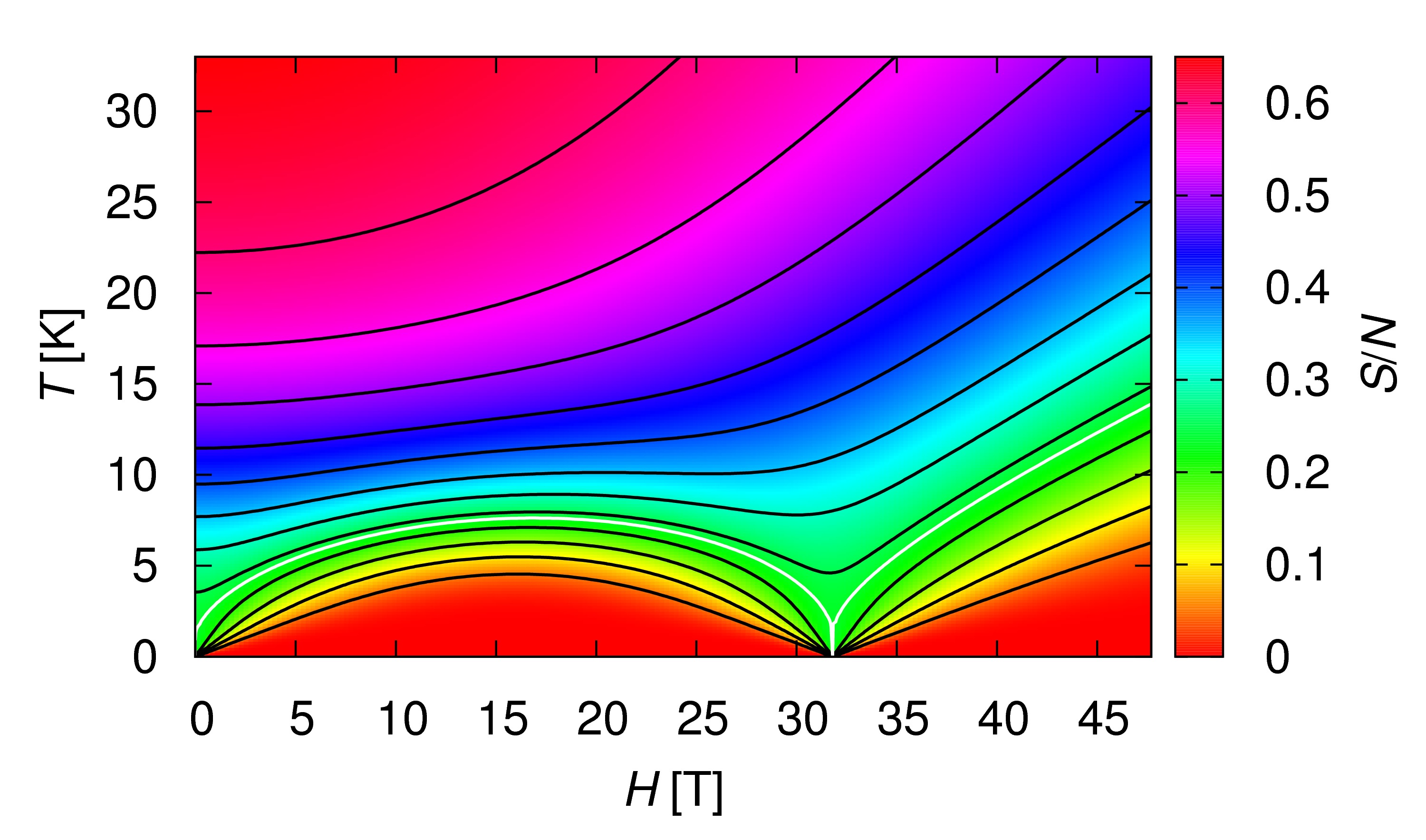} \\
\includegraphics[width=0.7\columnwidth]{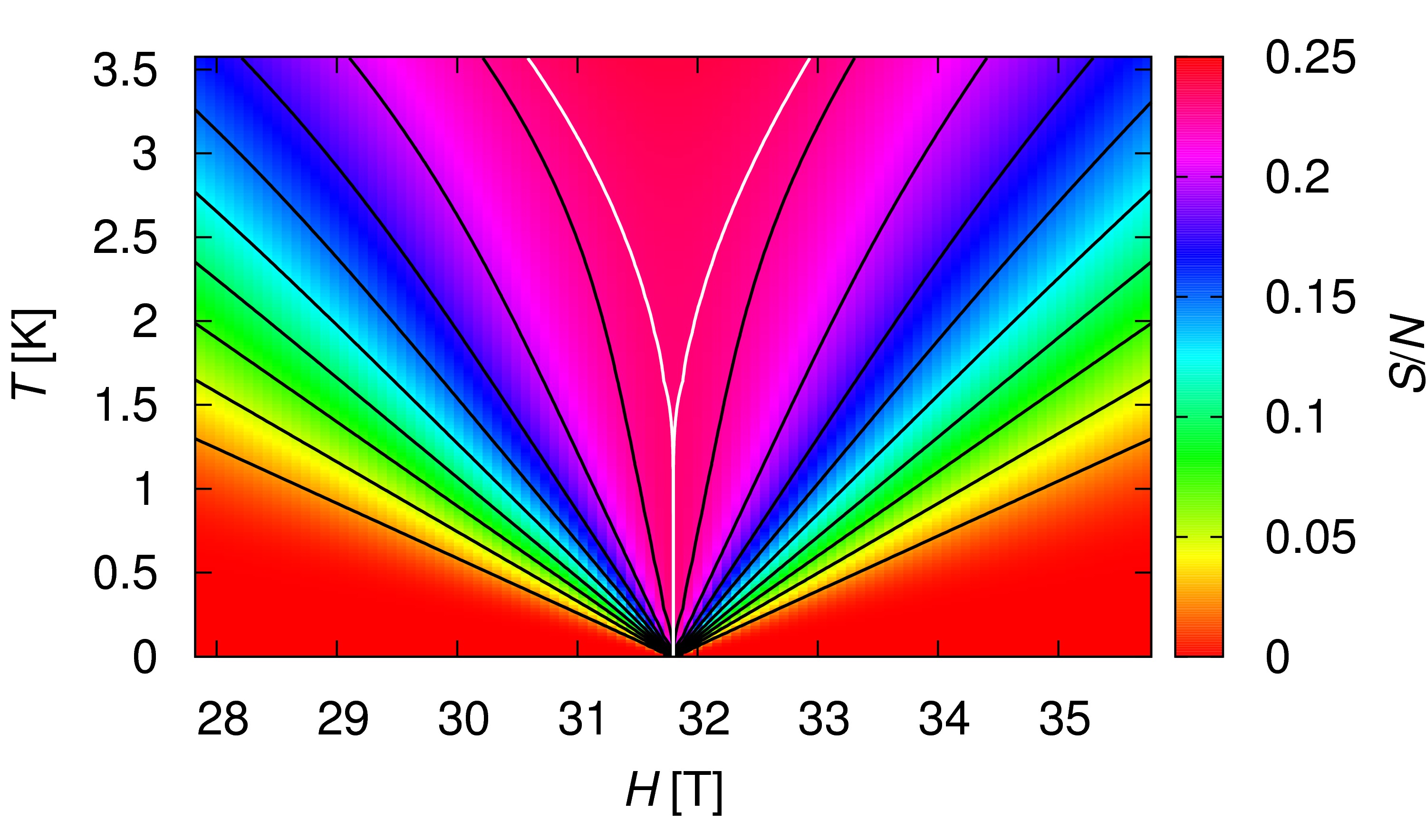}
\caption{
Entropy per spin $S/N$ of the ideal diamond chain
model with $J_1=J_3=11$K, $J_2=33$K and $J_{\rm m} = 0$,
as a function of magnetic field $H$ and temperature $T$.
The top panel covers the region from zero field to a fully polarised
system. In this panel the black lines correspond to
$S/N=0.05$, $0.1$, $0.15$, \ldots (in increasing order).
The bottom panel focuses on the low-temperature behaviour
at the transition to full polarisation. Here the black lines correspond to
$S/N=0.025$, $0.05$, $0.075$, \ldots The white lines in both
panels denote the residual entropy $S/N = (\ln 2)/3$. \\
The data in this figure has been obtained by exact diagonalisation
for $N=24$ spins.
\label{fig:entropy}
}
\end{center}
\end{figure}

\subsection{Ideal diamond chain}

Let us first take a look at an ideal diamond chain model with parameters
which are similar to those for azurite. Note that for $J_1=J_3$ one can
use conservation of the total spin on each vertical dimer
$\vec{S}_{{\rm d1},x} + \vec{S}_{{\rm d2},x}$ to speed up
the computation \cite{DRHS07}. Figure \ref{fig:entropy} shows the result
for $N=24$ spins and $J_1=J_3$, $J_2 = 3 \, J_1$. The large value of
$J_2$ ensures that all low-energy states are simple product states.
Hence, finite-size effects are negligible in figure \ref{fig:entropy}.

In a magnetic field the monomer spins are immediately polarised and thus
frozen. Exactly at the saturation field $H_{\rm sat.}$, the dimer singlet and
one component of the dimer triplet become degenerate, giving rise to
a two-fold degeneracy per dimer, i.e.\ a residual entropy $S/N=(\ln2)/3$
\cite{DRHS07,DeRi06}. On the other hand, for low magnetic fields the
dimers are frozen in their singlet state. Exactly at $H=0$, the
two projections of the monomer spins become degenerate and we
find again a residual entropy $S/N=(\ln2)/3$. This particular value
of the entropy is traced by the white lines in figure \ref{fig:entropy}.
Consider now an adiabatic process which is defined by a constant
entropy and thus follows the lines in  figure \ref{fig:entropy}.
If such an adiabatic process is started below the white lines
in figure \ref{fig:entropy}, i.e.\ with an entropy $S/N<(\ln2)/3$,
one achieves cooling to $T \to 0$ for $H \to H_{\rm sat.}$ (or for
$H \to 0$).

\begin{figure}[t]
\begin{center}
\includegraphics[width=0.7\columnwidth]{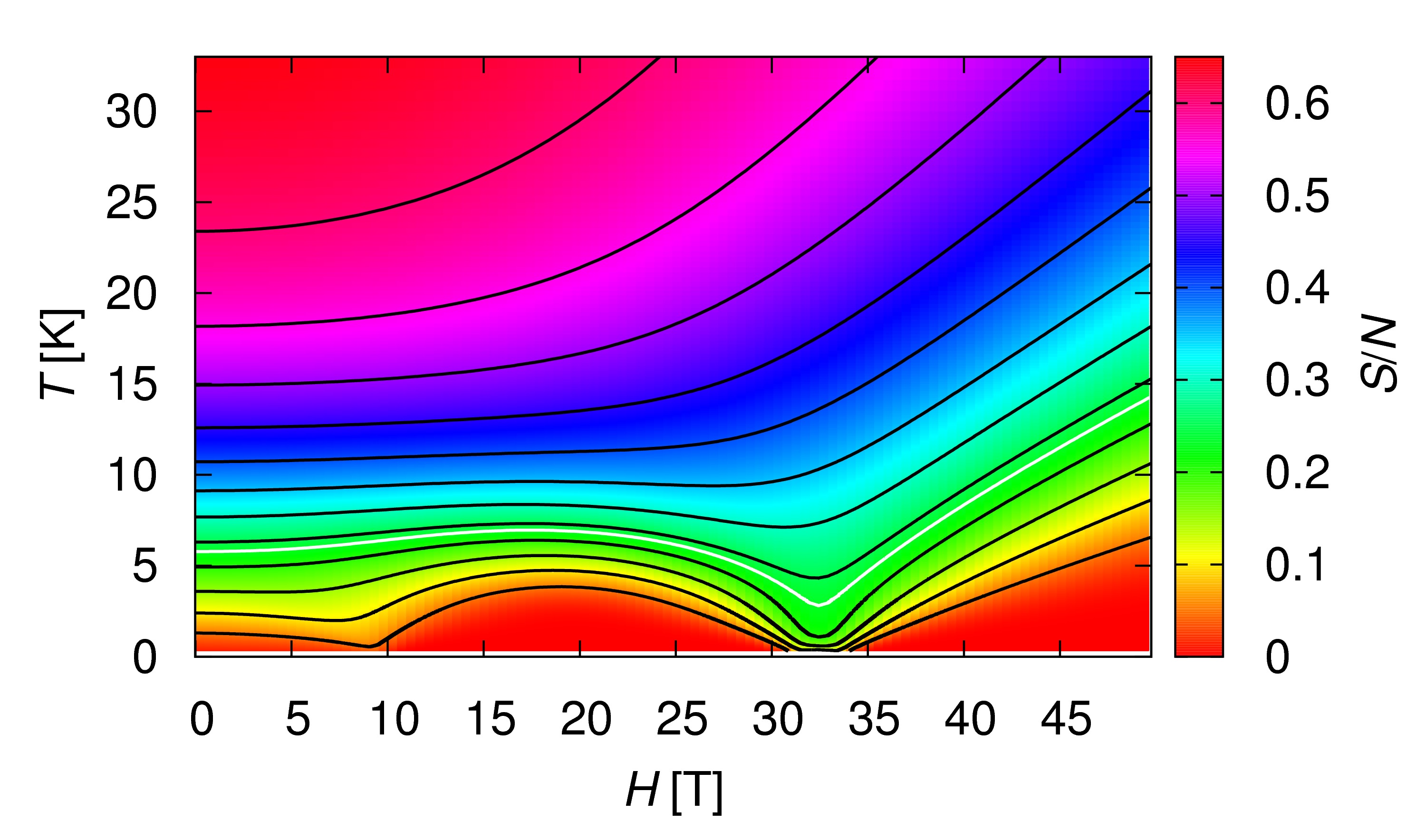} \\
\includegraphics[width=0.7\columnwidth]{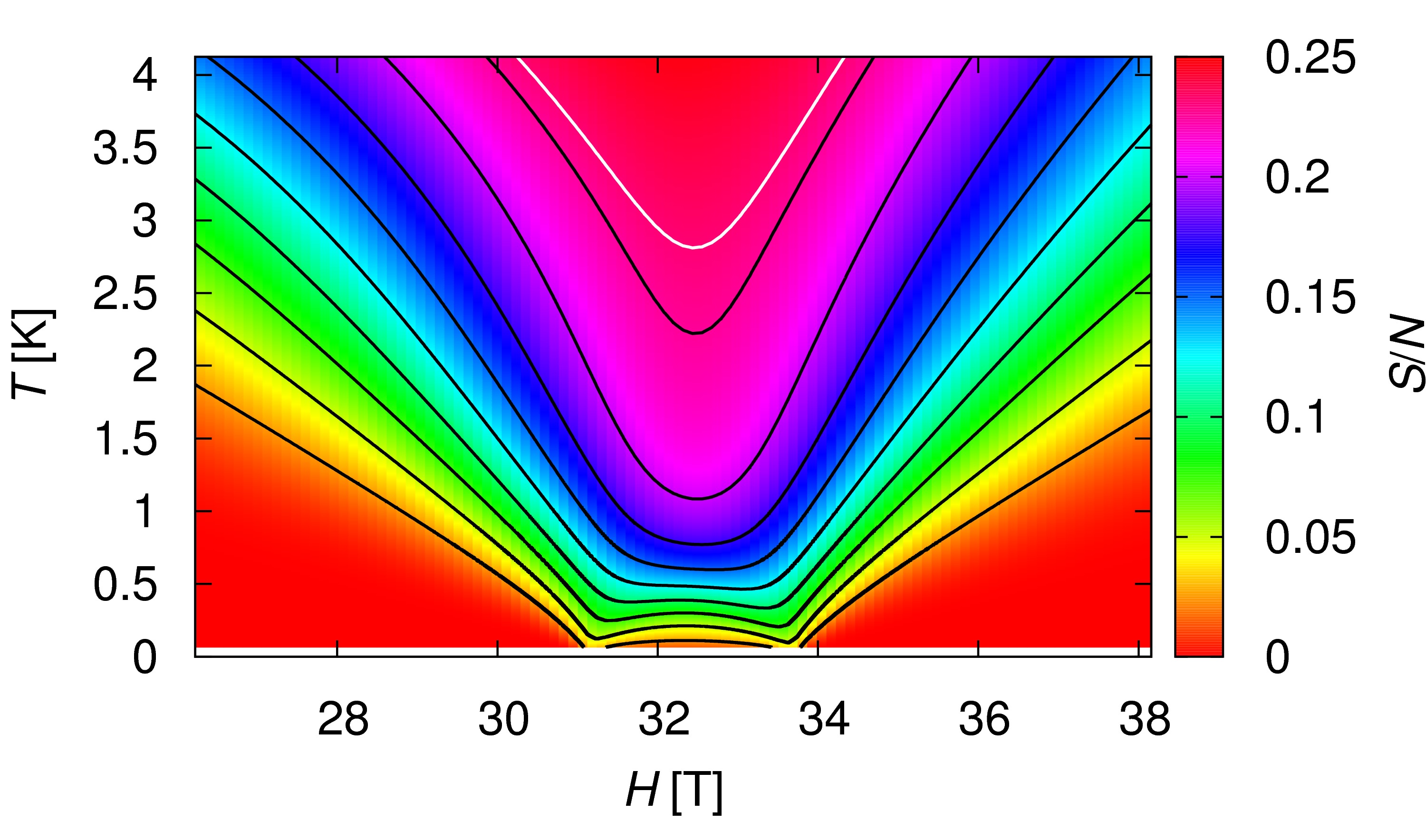}
\caption{
Entropy per spin $S/N$ of the generalised diamond chain model
for azurite,
i.e.\ $J_1=15.51$K, $J_2=33$K, $J_3=6.93$K and $J_{\rm m} = 4.62$K,
as a function of magnetic field $H$ and temperature $T$.
The top panel covers the region from zero field to a fully polarised
system. In this panel the black lines correspond to
$S/N=0.05$, $0.1$, $0.15$, \ldots (in increasing order).
The bottom panel focuses on the low-temperature behaviour
at the transition to full polarisation. Here the black lines correspond to
$S/N=0.025$, $0051$, $0.075$, \ldots The white lines in both
panels denote the residual entropy of the ideal diamond chain,
$S/N = (\ln 2)/3$. \\
The data in this figure is for the thermodynamic
limit and has been obtained by TMRG with $m=300$ kept states.
\label{fig:entropyGen}
}
\end{center}
\end{figure}

\subsection{Generalised diamond chain model for azurite}

Now we turn to the generalised diamond chain model for azurite
and check how $J_1 \ne J_3$ lifts the degeneracy. For this purpose we have
computed the entropy using a transfer-matrix variant of the
density-matrix renormalisation group, also known as `TMRG'
\cite{TMRG96,TMRG97}. Note that this method works for the infinite system
and proceeds at a fixed magnetic field $H$
from high to low temperatures by successively adding Trotter-Suzuki slices.
There is just one refinement which had to be implemented to get accurate results
at low temperatures and high magnetic fields. In this region, the
(asymmetric) reduced density matrix only has a small number of large
eigenvalues while all other eigenvalues are tiny. Therefore,
we use an additional reorthogonalisation procedure after the left
and right eigenvectors of the reduced density matrix are obtained by
exact diagonalisation. This allows us to keep more states and thus
improve accuracy. We have tested this procedure against exact results
for the entropy of the spin-1/2 Heisenberg chain \cite{exactS1o2}
and found excellent agreement.

Figure \ref{fig:entropyGen} shows the result for the entropy
of the generalised diamond chain with the parameters
(\ref{eq:Jazurite}).
In this case none of the constant entropy curves with $S/N>0$ goes
to $T=0$, reflecting the lifting of the degeneracy present
in the ideal diamond chain. In other words: in the present case the
entropy per site $S/N$ goes to zero as $T \to 0$ for {\it all values
of the magnetic field} $H$. 
The value of the residual entropy of the ideal diamond chain
$S/N = (\ln 2)/3$ is again shown by white lines in
figure \ref{fig:entropyGen}. One can read off
that this residual entropy is pushed up to $T>2.8$K.

The degeneracy at $H=0$ is particularly fragile. It is lifted not
only by a distortion $J_1 \ne J_3$, but also by a finite direct coupling
of the monomer spins $J_{\rm m} > 0$. This is reflected by a window
of $H_{c1} \approx 9.6$T for the polarisation of the monomer spins
\cite{JeschkeEtal} and the fact that the value $S/N = (\ln 2)/3$ is
pushed up to $T\approx 5.8$K for $H=0$ (compare the upper panel
of figure \ref{fig:entropyGen}).

On the other hand, the degeneracy at the saturation field would survive
a finite monomer-monomer coupling $J_{\rm m} > 0$ and is lifted only
by the distortion $J_1 \ne J_3$. Accordingly the region in which the
dimers are polarised is spread over a smaller field window of about
$\Delta H= 2.6$T width between $H_{c2}=31.1$T and $H_{\rm sat.}=33.7$T
\cite{JeschkeEtal} and
the value $S/N = (\ln 2)/3$ is attained already for $T\approx2.8$K
(see the lower panel of figure \ref{fig:entropyGen}). Although the entropy
$S/N = (\ln 2)/3$  is spread over a temperature window of approximately
$2.8$K, an adiabatic process can still cool to substantially lower temperatures in
this high-field region. For example, an adiabatic process which
starts at $(H,T) \approx (27{\rm T},1.6{\rm K})$ or $(38{\rm T},1.65{\rm K})$
would go down to $T < 70$mK as $H \to 31.1$T or $33.8$T,
respectively.  This case corresponds to an entropy $S/N=0.025$, i.e.\
the lowest curve in the bottom panel of \ref{fig:entropyGen}. For the
larger value $S/N=0.05$ (second curve in the lower panel of
figure  \ref{fig:entropyGen})
an adiabatic process would still cool to a minimum temperature
$T \approx 90$mK.

\begin{figure}[t]
\begin{center}
\includegraphics[width=0.7\columnwidth]{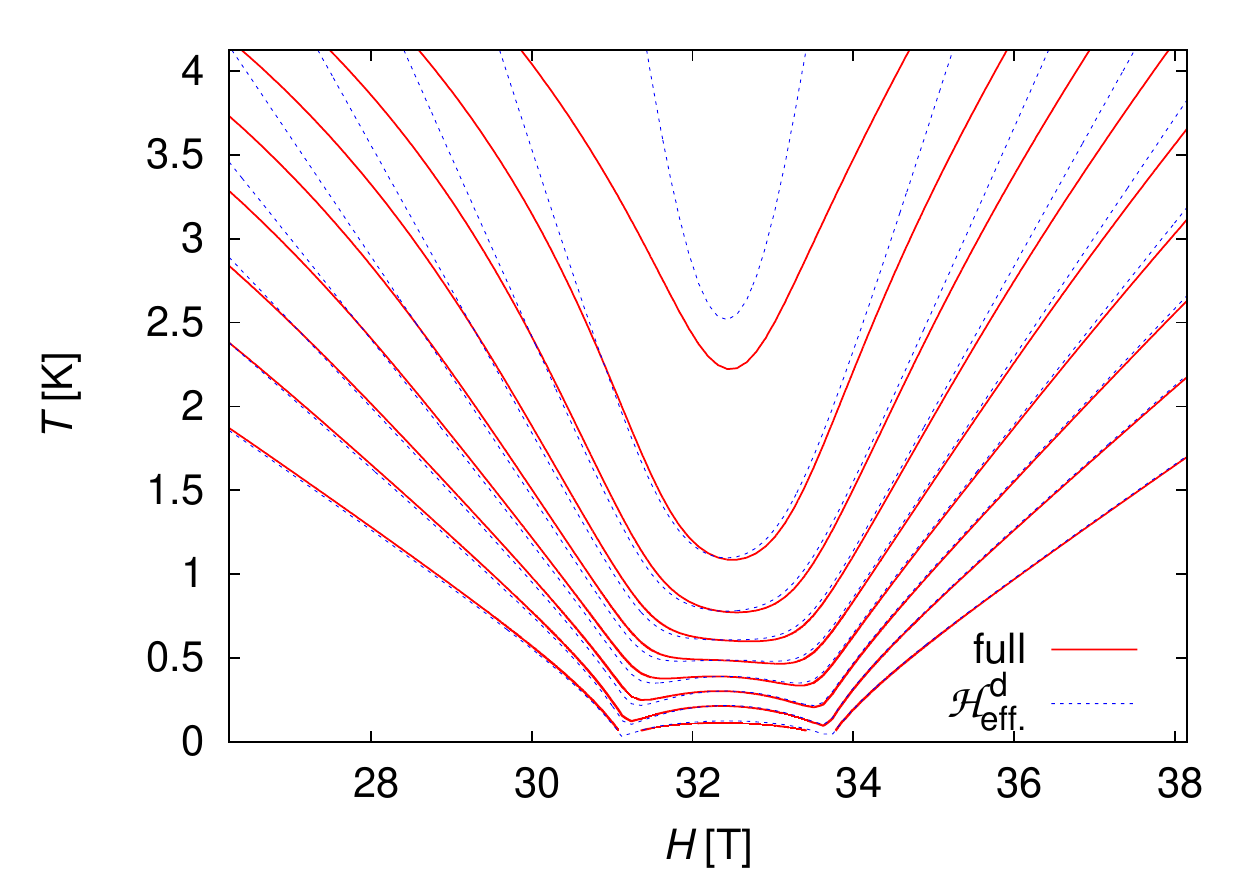}
\caption{Constant-entropy curves of the full generalised diamond chain model
(full lines) in comparison
to those of the effective Hamiltonian (\ref{eq:HeffD}) (dashed lines).
Lines correspond to $S/N=0.025$, $0051$, $0.075$, \ldots (in increasing order).
Results for the generalised diamond chain model correspond to the black 
constant entropy curves in the lower panel of figure \ref{fig:entropyGen}.
The entropy of the effective Hamiltonian (\ref{eq:HeffD}) has been
computed by exact diagonalisation with the parameters
(\ref{eq:JxyVal}), (\ref{eq:JzJdimerVal}) and $N/3=20$ dimer sites.
This system size is large enough to ensure the absence of visible finite-size
effects in the figure.
\label{fig:entropyComp}
}
\end{center}
\end{figure}

Finally, we compare the entropy in the high-field region shown in
the lower panel of figure \ref{fig:entropyGen} with the entropy of
the effective Hamiltonian (\ref{eq:HeffD}). The latter
can be considered as an effective description of the splitting of the
manifold of the $2^{N/3}$ states which are degenerate in the ideal
diamond chain exactly at $H_{\rm sat.}$.

In principle, the entropy could be computed exactly for the model
(\ref{eq:HeffD}) \cite{exactS1o2}. However,
we found it simpler to perform a full diagonalisation of this
model with the parameters (\ref{eq:JxyVal}), (\ref{eq:JzJdimerVal})
and $N/3=20$ dimer sites. This system size is large enough to ensure
the absence of visible finite-size effects in the dashed constant-entropy
curves in figure \ref{fig:entropyComp}.

The constant entropy curves of the full model which are shown by black
lines in the lower panel of figure \ref{fig:entropyGen} are reproduced
in figure \ref{fig:entropyComp} by full lines. The dashed lines show
results for the effective Hamiltonian (\ref{eq:HeffD}) with the same
values of the entropy $S/N$. One finds that the effective model reproduces
the results of the full one well for temperatures $T \lesssim 1.5$K
and in the high-field regime shown in figure \ref{fig:entropyComp}.
The deviations observed at higher temperatures in
figure \ref{fig:entropyComp} indicate that other states
become relevant in this temperature range. Note that the uppermost
curve in figure \ref{fig:entropyComp} corresponds to $S/N = 0.225$
which is very close to the total entropy $S/N = (\ln 2)/3$ of the
effective chain model.
Still, a simplified description in terms of the effective Hamiltonian
(\ref{eq:HeffD}) works well in high magnetic fields and at sufficiently
low temperatures.

\section{Summary and discussion}

\label{sec:SumDisc}

In this work
we have explored further properties of the generalised diamond chain
model for azurite \cite{JeschkeEtal}. First, we have computed the excitation
spectrum in the absence of a magnetic field and found a two-spinon
continuum at low energies as well as a dimer branch sitting on a
broad background at higher energies. These features are in good agreement
with inelastic neutron scattering results on azurite \cite{azuritINS},
thus further supporting the description of azurite in terms of a
generalised diamond chain model \cite{JeschkeEtal}.

We have then computed magnetocaloric properties at high magnetic fields.
The degeneracy present in the ideal model at the saturation field
\cite{DRHS07,DeRi06} is lifted in the generalised diamond chain
model for azurite. Still, we predict cooling capabilities down to temperatures
substantially below $1$K as the magnetic field approaches the upper edge of
the one-third plateau $H_{c2}$ or the saturation field $H_{\rm sat.}$
from below or above, respectively.

There are at least two features in azurite which are not accounted for
by the generalised diamond chain model. Firstly, a magnetic anisotropy
is clearly present in azurite \cite{KikuchiA}. This is most likely due
to Dzyaloshinsky-Moriya interactions. Indeed, there is at least one
investigation of the effect of such terms \cite{SOT10}, however not for
the parameters which we consider to be most appropriate for azurite.
On the other hand, the complication of magnetic anisotropies can be
experimentally avoided by aligning the external magnetic field with
the high-symmetry axis.

Secondly, azurite orders for temperatures below $2$K in low magnetic fields
\cite{KikuchiA,SpE58,Forstat,gibson10,rule10}. This ordering process reflects
the presence of interchain coupling terms \cite{Whangbo,JeschkeEtal} and
could account in particular for the fact that we predict the dimer branch
at zero magnetic field at energies which are a few meV below the
experimental result \cite{azuritINS}\footnote{Interchain
coupling and magnetic anisotropies can also give rise to additional
finer structures in the spectra which are visible in high-resolution
inelastic neutron scattering experiments \cite{RulePrivCom11}.}.
Also in high magnetic fields
azurite is found to be ordered at $T=600$mK \cite{HorvaticPrivCom10}.
Such an ordering transition will push most of the low-temperature
entropy up to the transition temperature such that most of the
adiabatic (de)magnetisation curves are pushed to temperatures
higher than $600$mK. Nevertheless, the ordering temperature is
expected to vanish as one approaches the one-third plateau and the
fully polarised state at $H=H_{c2}$ and $H_{\rm sat.}$, respectively.
Thus, we expect
the strong cooling effect at $H_{c2}$ and $H_{\rm sat.}$ to be preserved
in azurite
as these two fields are approached from below and above, respectively.

A more accurate treatment of the interchain coupling geometry
\cite{JeschkeEtal} is clearly necessary for a quantitative
description of the ordered states of azurite at temperatures
below $2$K. In this context, it may be interesting to note that we can
obtain very accurate results using effective Hamiltonians, provided that
their parameters are determined carefully. On the one hand, we have
shown that the effective Heisenberg chain (\ref{eq:HeffM}) yields a
very good description of the low-energy excitations at $H=0$
(see figure \ref{fig:specH0}). On the other hand,
figure \ref{fig:entropyComp} demonstrates that the low-temperature
behaviour of the generalised diamond chain in high magnetic fields
is well described by the effective Hamiltonian (\ref{eq:HeffD}).
The high frustration of the underlying model is reflected by the
small values of the effective exchange constants $J_{\rm xy}$
and $J_{\rm z}$ which are one to two orders of magnitude smaller than the bare
exchange constants.

In particular, the experimental observation
of simple antiferromagnetic order in azurite
at high magnetic fields \cite{HorvaticPrivCom10}
can be rationalised by noting that the parameters (\ref{eq:JxyVal}) and
(\ref{eq:JzJdimerVal}) lead to an effective easy-plane anisotropy. Under such
conditions an instability towards antiferromagnetic order in the transverse
components is expected. However, for a quantitative description of
the low-energy properties of azurite in the high-field region, interchain
coupling definitely needs to be included in an effective Hamiltonian,
in particular in view of the fact that the effective one-dimensional
exchange constant (\ref{eq:JxyVal}) is comparable to the maximal ordering
temperature in the high-field regime \cite{HorvaticPrivCom10}.

Similar effective Hamiltonians have been successfully employed, e.g.,
in the context of SrCu$_2$(BO$_3$)$_2$ \cite{DSM08}. However, the
model for azurite \cite{JeschkeEtal} has the advantage that the bare dimer
exchange constant $J_2$ is substantially bigger than any other
exchange constant. Hence, we may not only expect less terms to be relevant in
the effective Hamiltonians than in the case of SrCu$_2$(BO$_3$)$_2$
\cite{DSM08}, but also better quantitative validity
for the parameters appropriate to azurite \cite{JeschkeEtal}
if the parameters of the effective models are determined carefully.

Finally, we note that
the magnetocaloric effect has been widely used at a qualitative
level for the experimental
determination of the phase diagram of spin systems in a magnetic
field (see, e.g., \cite{JCNBKKJSHZSU04,SJSBJBF08,RKTMZNZBKGCPBK08,
FHYSOTT09,SKDSHBJF09,AKJNCBDL09,AKMWJVMSDL09}) and there is indirect
evidence for a strong magnetocaloric effect in SrCu$_2$(BO$_3$)$_2$
\cite{LSBHTKWU08}. However, as far as we are aware, there are only
very few quantitative measurements of the cooling capabilities of
quantum spin systems \cite{LTWJTHRPAD10,WTJTHRHPADL} and in particular
highly frustrated magnets \cite{SPSGBPBZ05,RTCGTS07}. Measurements
of the magnetocaloric effect in the high-field region of
azurite would therefore certainly be very interesting.

\ack
We are grateful to M.\ Horvati\'c and K.\ Rule for useful discussions
and comments.
We acknowledge allocation of the CPU time on the High
Performance Computer Cluster ``Kohn'' in the Department of Physics, Renmin
University of China as well as the Theory/Grid Cluster at G\"ottingen
University.
A.H.\ and J.R.\ would like to thank the DFG for financial support via
a Heisenberg fellowship under project HO~2325/4-2 and under
project RI615/16-1, respectively.
S.H.\ would like to acknowledge support via a LiSUM fellowship
as well as by grants NSFC10874244 and MSTC2007CB925001.
R.P.\ is supported by the Japan Society for the Promotion of Science (JSPS)
and the Alexander von Humboldt-Foundation.


\section*{References}

\begin{thebibliography}{19}

\bibitem{HFMbook} Lacroix C, Mendels P and Mila F 2011
 Introduction to Frustrated Magnetism
 ({\it Springer Series in Solid-State Sciences} vol.\ 164)
 (Heidelberg: Springer)

\bibitem{TrWi05} Troyer M and Wiese U-J 2005
 \PRL {\bf 94} 170201

\bibitem{MiUe03} Miyahara S and Ueda K 2003
 \JPCM {\bf 15} R327

\bibitem{SSRS01} Schnack J, Schmidt H-J, Richter J and Schulenburg J 2001
 {\it  Eur. Phys. J.} B {\bf 24} 475

\bibitem{loc_mag02} Schulenburg J, Honecker A, Schnack J, Richter J and
     Schmidt H-J 2002
     \PRL {\bf 88} 167207

\bibitem{LNP04} Richter J, Schulenburg J and Honecker A 2004
 {\it Lect. Notes Phys.} {\bf 645} 85

\bibitem{ZhHo04} Zhitomirsky ME and Honecker A 2004
  {\it J. Stat. Mech.: Theor. Exp.} P07012

\bibitem{ZhiTsu04} Zhitomirsky ME and Tsunetsugu H 2004
  {\it Phys. Rev.} B {\bf 70} 100403(R)

\bibitem{DR04} Derzhko O and Richter J 2004
  {\it Phys. Rev.} B {\bf 70} 104415

\bibitem{RSHS04} Richter J, Schulenburg J, Honecker A and Schmalfu{\ss} D
  2004
  {\it Phys. Rev.} B {\bf 70} 174454

\bibitem{SP2004} Richter J, Derzhko O and  Schulenburg J 2004
       {\it Phys. Rev. Lett.} {\bf 93} 107206 

\bibitem{loc_mag04}
     Richter J, Schulenburg J, Honecker A, Schnack J and Schmidt H-J 2004
     \JPCM {\bf 16} S779

\bibitem{ZhiTsu05} Zhitomirsky ME and Tsunetsugu H 2005
 {\it Prog. Theor. Phys. Suppl.} {\bf 160} 36

\bibitem{SRS05} Schmidt R, Richter J and Schnack J 2005
  {\it J. Magn. Magn. Mater.} {\bf 295} 164

\bibitem{DeRi06} Derzhko O and Richter J 2006
  {\it Eur. Phys. J.} B {\bf 52} 23

\bibitem{SSHSR06} Schnack J, Schmidt H-J, Honecker A, Schulenburg J and
  Richter J 2006
  {\it J. Phys.: Conf. Ser.} {\bf 51} 43

\bibitem{SRS07} Schnack J, Richter J and Schmidt R 2007
  {\it Phys. Rev.} B {\bf 76} 054413

\bibitem{ZT07} Zhitomirsky ME and Tsunetsugu H 2007
  {\it Phys. Rev.} B {\bf 75} 224416

\bibitem{DRHS07} Derzhko O, Richter J, Honecker A and Schmidt H-J 2007
  {\it Low Temp. Phys.} {\bf 33} 745

\bibitem{RLM08} Rousochatzakis I, L\"auchli AM and Mila F 2008
  {\it Phys. Rev.} B {\bf 77} 094420

\bibitem{RDH08} Richter J, Derzhko O and Honecker A 2008
  {\it Int. Jour. Mod. Phys. B} {\bf 22} 4418

\bibitem{Schnack10} Schnack J 2010
  {\it Dalton Trans.} {\bf 39} 467


\bibitem{Zhito03} Zhitomirsky ME 2003
 {\it Phys. Rev.} B {\bf 67} 104421

\bibitem{SPSGBPBZ05} Sosin SS, Prozorova LA, Smirnov AI
  Golov AI, Berkutov IB, Petrenko OA, Balakrishnan G and Zhitomirsky ME 2005
  {\it Phys. Rev.} B {\bf 71} 094413

\bibitem{ZhHo09} Honecker A and Zhitomirsky ME 2009
  {\it J. Phys.: Conf. Ser.} {\bf 145} 012082 

\bibitem{PML09} Pereira MSS, de Moura FABF and Lyra ML 2009
  {\it Phys. Rev.} B {\bf 79} 054427


\bibitem{Gelfand91} Gelfand MP 1991
 {\it Phys. Rev.} B {\bf 43} 8644

\bibitem{IvRi97} Ivanov NB and Richter J 1997
 {\it Phys. Lett.} A {\bf 232} 308

\bibitem{RIS98}  Richter J, Ivanov NB and Schulenburg J 1998
 \JPCM {\bf 10} 3635

\bibitem{MTM99} Mambrini M, Tr\'ebosc J and Mila F 1999
 {\it Phys. Rev.} B {\bf 59} 13806 

\bibitem{KOK00}  Koga A, Okunishi K and Kawakami N 2000
 {\it Phys. Rev.} B {\bf 62}  5558

\bibitem{HMT00} Honecker A,  Mila F and Troyer M,
 {\it Eur. Phys. J.} B {\bf 15} 227

\bibitem{MHSKU00} M\"uller-Hartmann E, Singh RRP, Knetter C and Uhrig GS 2000
 {\it Phys. Rev. Lett.} {\bf 84} 1808
 
\bibitem{GVAHW00} Gros C, Valent\'{\i} R, Alvarez JV, Hamacher K
  and Wenzel W 2000
  {\it Phys. Rev.} B {\bf 62} R14617

\bibitem{TrSe00} Trebst S and Sengupta S 2000
  {\it Phys. Rev.} B {\bf 62} R14613

\bibitem{HoBr01} Honecker A and Brenig W 2001
  {\it Phys. Rev.} B {\bf 63} 144416

\bibitem{ChBo02} Chattopadhyay E and Bose I 2002
  {\it Phys. Rev.} B {\bf 65} 134425

\bibitem{SchRi02a} Schulenburg J and Richter J 2002
  {\it Phys. Rev. B} {\bf 65} 054420

\bibitem{SchRi02b} Schulenburg J and Richter J 2002
  {\it Phys. Rev.} B {\bf 66} 134419

\bibitem{ChBue02} Chen S and B\"uttner H 2002
  {\it Eur. Phys. J.} B {\bf 29} 15

\bibitem{RoAl03} Rojas O and Alcaraz F C 2003
  \PR B {\bf 67} 174401

\bibitem{RDT06} Richter J, Derzhko O and Krokhmalskii T 2006
  {\it Phys. Rev.} B {\bf 74} 144430

\bibitem{BWB08} Bergman DL, Wu C and Balents L 2008
   {\it Phys. Rev.} B {\bf 78} 125104

\bibitem{MPSM10} Manmana SR, Picon J-P, Schmidt KP and Mila F 2010
 Unconventional magnetization plateaus in a Shastry-Sutherland spin tube
 {\it Preprint} arXiv:1003.1696

\bibitem{DTK10} Derzhko O, Krokhmalskii T and Richter J 2010
 \PR B {\bf 82} (2010) 214412


\bibitem{TKS96} Takano K, Kubo K and Sakamoto H 1996
 \JPCM {\bf 8} 6405

\bibitem{strechka1}
\v{C}anova L, Stre\v{c}ka J and  Ja\v{s}\v{c}ur M 2006
\JPCM  {\bf 18} 4967

\bibitem{strechka2} \v{C}anova L, Stre\v{c}ka J and  Lu\v{c}ivjansk\'y T 2009
 {\it Condensed Matter Physics} {\bf 12} 353

\bibitem{ivanov}
Ivanov NB, Richter J and Schulenburg J 2009
 {\it Phys. Rev.} B {\bf 79} 104412 

\bibitem{hida1} Hida K, Takano K and Suzuki H 2010
 \JPSJ {\bf 79} 044702 

\bibitem{hida2} Hida K, Takano K and Suzuki H 2010
 \JPSJ {\bf 79} 114703 

\bibitem{rsok10}  Rojas O, de Souza SM, Ohanyan V and Khurshudyan M 2010
Exactly solvable model of Ising-Heisenberg diamond-chain with $S=1$ $XXZ$
vertical dimers with additional biquadratic interactions and single-ion anisotropy
 {\it Preprint} arXiv:1007.0098


\bibitem{OTTK99} Okamoto K, Tonegawa T, Takahashi Y and Kaburagi M 1999
 \JPCM {\bf 11} 10485

\bibitem{TOHTK00} Tonegawa T, Okamoto K, Hikihara T, Takahashi Y and
  Kaburagi M 2000
  \JPSJ {\bf 69} Suppl. A 332

\bibitem{TOHTK01} Tonegawa T, Okamoto K, Hikihara T, Takahashi Y and
  Kaburagi M 2001
 {\it J. Phys. Chem. Solids} {\bf 62} 125

\bibitem{HoL} Honecker A and L\"auchli A 2001
   {\it Phys. Rev.} B {\bf 63} 174407

\bibitem{OTK03} Okamoto K, Tonegawa T and Kaburagi M 2003
  \JPCM {\bf 15} 5979

\bibitem{SOT09} Sakai T, Okamoto K and Tonegawa T 2009
   {\it J. Phys.: Conf. Ser.} {\bf 145} 012065


\bibitem{OYA97} Oshikawa M, Yamanaka M and Affleck I 1997
  \PRL {\bf 78} 1984

\bibitem{CHP97} Cabra DC, Honecker A and Pujol P 1997
  \PRL {\bf 79} 5126

\bibitem{CHP98} Cabra DC, Honecker A and Pujol P 1998
  {\it Phys. Rev.} B {\bf 58} 6241

\bibitem{KikuchiA} Kikuchi H, Fujii Y, Chiba M, Mitsudo S,
  Idehara T, Tonegawa T, Okamoto K, Sakai T, Kuwai T and
  Ohta H 2005
  {\it Phys. Rev. Lett.} {\bf 94} 227201

\bibitem{KikuchiB} Kikuchi H, Fujii Y, Chiba M, Mitsudo S,
  Idehara T, Tonegawa T, Okamoto K, Sakai T, Kuwai T and Ohta H 2006
  \PRL {\bf 97} 089702

\bibitem{MiLu} Mikeska H-J and Luckmann C 2008
  {\it Phys. Rev.} B {\bf 77} 054405

\bibitem{GuSu1} Gu B and Su G 2006
  \PRL {\bf 97} 089701

\bibitem{GuSu2} Gu B and Su G 2007
  {\it Phys. Rev.} B {\bf 75} 174437

\bibitem{azuritINS} Rule KC, Wolter AUB, S\"ullow S,
  Tennant DA, Br\"uhl A, K\"ohler S, Wolf B, Lang M and
  Schreuer J 2008
 {\it Phys. Rev. Lett.} {\bf 100} 117202

\bibitem{Whangbo} Kang J, Lee C, Kremer RK and Whangbo M-H
 2009
  \JPCM {\bf 21} 392201

\bibitem{JeschkeEtal} Jeschke H, Opahle I, Kandpal H, Valent\'{\i} R,
  Das H, Saha-Dasgupta T, Janson O, Rosner H, Br\"uhl A, Wolf B,
  Lang M, Richter J, Hu S, Wang X, Peters R, Pruschke T and
  Honecker A  2010
  Multi-step approach to microscopic models for frustrated quantum
  magnets -- the case of the mineral azurite
  {\it Preprint} arXiv:1012.2269

\bibitem{DMRGa} White SR 1992
  \PRL {\bf 69} 2863

\bibitem{DMRGb} Schollw\"ock U 2005
  {\it Rev. Mod. Phys.} {\bf 77} 259

\bibitem{KikuchiE} Kikuchi H, Fujii Y, Chiba M, Mitsudo S,
  Idehara T, Tonegawa T, Okamoto K, Sakai T, Kuwai T, Kindo K,
  Matsuo A, Higemoto W, Nishiyama K, Horvati\'c M and Berthier C 2005
  {\it Progr. Theor. Phys. Suppl.} {\bf 159} 1

\bibitem{azuritNMR} Aimo F, Kr\"amer S, Klanj\v{s}ek M,
  Horvati\'c M, Berthier C and Kikuchi H 2009
  \PRL {\bf 102} 127205

\bibitem{Ohta} Ohta H, Okubo, S, Kamikawa T, Kunimoto T, Inagaki Y,
              Kikuchi H, Saito T, Azuma M and Takano M 2003
             \JPSJ {\bf 72} 2464

\bibitem{dDMRG} Jeckelmann E 2002
  {\it Phys. Rev.} B {\bf 66} 045114

\bibitem{TMRG96} Bursill R~J, Xiang T and Gehring GA 1996
 \JPCM {\bf 8} L583

\bibitem{TMRG97} Wang X and Xiang T 1997
  {\it Phys. Rev.} B {\bf 56} 5061

\bibitem{MTBB81} M\"uller G, Thomas H, Beck H and Bonner JC 1981 
 \PR B {\bf 24} 1429

\bibitem{HoWe09} Honecker A and Wessel S 2009
  {\it Condensed Matter Physics} {\bf 12} 399

\bibitem{FuGue79} Furrer A and G\"udel H-U 1979
  {\it J. Magn. Magn. Mater.} {\bf 14} 256

\bibitem{Zigan:72} Zigan F and Schuster HD 1972
  {\it Z. Kristallogr.} {\bf 135} 416.

\bibitem{ClPe62} des Cloizeaux J and Pearson JJ 1962
 {\it Phys. Rev.} {\bf 128} 2131 (1962) 

\bibitem{FaTa81} Faddeev LD and Takhtajan LA 1981
 {\it Phys. Lett.} A {\bf 85} 375

\bibitem{LTFN05} Lake B, Tennant DA, Frost CD and Nagler SE 2005
 {\it Nature Materials} {\bf 4} 329

\bibitem{CaHa06} Caux J-S and Hagemans R 2006
 {\it J. Stat. Mech.: Theor. Exp.} P12013

\bibitem{exactS1o2} Trippe C, Honecker A, Kl\"umper A and Ohanyan V 2010 
           {\it Phys. Rev.} B {\bf 81} 054402

\bibitem{SOT10} Sakai T, Okamoto K and Tonegawa T 2010
 {\it J. Phys.: Conf. Ser.} {\bf 200} 022052

\bibitem{SpE58}  Spence RD and Ewing RD 1958
 \PR {\bf 112} 1544

\bibitem{Forstat} Forstat H, Taylor G and King RB 1959
  {\it J. Chem. Phys.} {\bf 31} 929

\bibitem{gibson10} Gibson MCR, Rule KC, Wolter AUB,
  Hoffmann J-U, Prokhnenko O, Tennant DA, Gerischer S,
  Kraken M, Litterst FJ, S\"ullow S, Schreuer J, Luetkens
  H, Br\"uhl A, Wolf B and Lang M 2010
  \PR B  {\bf 81} 140406(R)

\bibitem{rule10} Rule KC, Reehuis M, Gibson MCR, Ouladdiaf B, Gutmann MJ,
  Hoffmann J-U, Gerischer S, Tennant DA, S\"ullow S and  Lang M 2011
  \PR B {\bf 83} 104401

\bibitem{RulePrivCom11} Rule KC 2011
  {\it private communication}

\bibitem{HorvaticPrivCom10} Aimo F, Kr\"amer S, Klanj\v{s}ek, Horvati\'c M
  and Berthier C 2011
  Magnetic structure of azurite above the 1/3 magnetization plateau
  {\it Preprint} arXiv:1103.0376

\bibitem{DSM08} Dorier J, Schmidt KP and Mila F 2008
  {\it Phys. Rev. Lett.} {\bf 101} 250402


\bibitem{JCNBKKJSHZSU04} Jaime M, Correa VF, N. Harrison, Batista CD,
 Kawashima N, Kazuma Y, Jorge GA, Stern R, Heinmaa I, Zvyagin SA,
 Sasago Y and Uchinokura K 2004
 \PRL {\bf 93} 087203

\bibitem{SJSBJBF08} Samulon EC, Jo Y-J, Sengupta P, Batista CD,
 Jaime M, Balicas L and Fisher IR 2008
 \PR B {\bf 77} 214441

\bibitem{RKTMZNZBKGCPBK08} R\"uegg Ch, Kiefer K, Thielemann B,
  McMorrow DF, Zapf V, Normand B, Zvonarev MB, Bouillot P,
  Kollath C, Giamarchi T, Capponi S, Poilblanc D, Biner D and Kr\"amer KW 2008
 \PRL {\bf 101} 247202

\bibitem{FHYSOTT09} Fortune NE, Hannahs ST, Yoshida Y, Sherline TE,
 Ono T, Tanaka H and Takano Y 2009
 \PRL {\bf 102} 257201

\bibitem{SKDSHBJF09} Samulon EC, Kohama Y, McDonald RD, Shapiro MC,
 Al-Hassanieh KA, Batista CD, Jaime M and Fisher IR 2009
 \PRL {\bf 103} 047202

\bibitem{AKJNCBDL09} Aczel AA, Kohama Y, Jaime M, Ninios K,
 Chan HB, Balicas L, Dabkowska HA and Luke GM 2009
 \PR B {\bf 79} 100409(R)

\bibitem{AKMWJVMSDL09} Aczel AA, Kohama Y, Marcenat C, Weickert F, Jaime M,
 Ayala-Valenzuela OE, McDonald RD, Selesnic SD, Dabkowska HA and Luke GM 2009
 \PRL {\bf 103} 207203

\bibitem{LSBHTKWU08}  Levy F, Sheikin I, Berthier C, Horvati\'c M,
  Takigawa M, Kageyama H, Waki T and Ueda Y 2008
  {\it Europhys. Lett.} {\bf 81} 67004

\bibitem{LTWJTHRPAD10} Lang M, Tsui Y, Wolf B, Jaiswal-Nagar D, Tutsch U,
 Honecker A, Removi\'c-Langer K, Prokofiev A, Assmus W and Donath G 2010
  {\it J. Low Temp. Phys.} {\bf 159} 88

\bibitem{WTJTHRHPADL} Wolf B, Tsui Y, Jaiswal-Nagar D, Tutsch U,
 Honecker A, Removi\'c-Langer K, Hofmann G, Prokofiev A, Assmus W,
 Donath G and Lang M 2010
 Magnetocaloric effect and magnetic cooling near a field-induced quantum-critical point
{\it Preprint} arXiv:1012.3328

\bibitem{RTCGTS07} Radu T, Tokiwa Y, Coldea R, Gegenwart P, Tylczynski Z
  and Steglich F 2007
 {\it Sci. Technol. Adv. Mater.} {\bf 8} 406


\end{thebibliography}
\end{document}